# Realization of a High Mobility Dual-gated Graphene Field Effect Transistor with Al$_2$O$_3$ Dielectric

Seyoung Kim[1], Junghyo Nah[1], Insun Jo[2], Davood Shahrjerdi[1], Luigi Colombo[3], Zhen Yao[2], Emanuel Tutuc[1] and Sanjay K. Banerjee[1]

[1]*Microelectronics Research Center, Department of Electrical and Computer Engineering, The University of Texas at Austin, Austin, TX 78758, USA*
[2]*Department of Physics, The University of Texas at Austin, Austin, TX 78712, USA*
[3]*Texas Instruments, Inc. 12500 TI Boulevard Dallas, TX 75266, USA*

We fabricate and characterize dual-gated graphene field-effect transistors (FETs) using Al$_2$O$_3$ as top-gate dielectric. We use a thin Al film as a nucleation layer to enable the atomic layer deposition of Al$_2$O$_3$. Our devices show mobility values of over 8,000 cm$^2$/Vs at room temperature, a finding which indicates that the top-gate stack does not significantly increase the carrier scattering, and consequently degrade the device characteristics. We propose a device model to fit the experimental data using a single mobility value.

Graphene, a mono- to few-layers of *sp$^2$* bonded carbon in a honeycomb lattice, has been studied intensively since its discovery in 2004[1] due to its unique electron physics, as well as possible applications to electronic devices. Graphene's high intrinsic carrier mobility (over 200,000 cm$^2$/Vs at low temperature for suspended samples[2]), combined with its mechanical and thermodynamic stability[3], makes it a promising material for nano-electronic devices.

The fabrication of graphene-based field-effect transistors requires a uniform gate dielectric deposition technique on graphene, with high dielectric constant (κ) and reduced interface states density. It is well known that the existence of a mechanically and chemically stable native oxide for silicon, SiO$_2$, has been key to the success of silicon-based microelectronics. Highly insulating SiO$_2$ grows on Si by thermal oxidation[4], and the interface between Si and SiO$_2$ has almost close-to-ideal properties.[5] Atomic layer deposition (ALD) is a well developed technique used for growing high-κ gate dielectric layers, thanks to its precise control over the film thickness and uniformity.[6] However, the direct deposition of high-κ dielectric materials, such as Al$_2$O$_3$ and HfO$_2$, on graphene using H$_2$O-based ALD is not possible because of the hydrophobic nature of graphene basal plane.[7] Given that a perfect graphite surface is chemically inert[8], attempts to grow ALD Al$_2$O$_3$ layer on a *clean* HOPG (highly oriented pyrolytic graphite) surface lead to a selective growth at the steps between graphite layers, where the broken carbon bonds along the terraces serve as 1D nucleation center for the initial ALD process.[9] Therefore, the deposition of high-κ dielectric materials on graphene has been relatively limited so far.

Previous studies have used surface treatments of the graphene surface in order to allow ALD growth. Examples include NO$_2$ functionalization[10], O$_3$ functionalization[7], and PTCA (perylene tetracarboxylic acid) coating[11], or simply nucleating the dielectric growth from impurities on graphene without prior cleaning.[12] The carrier mobility on top-gated graphene devices with Al$_2$O$_3$ dielectric deposited using NO$_2$ functionalization was 7,000 cm$^2$/Vs at 4.2K.[10] On the other hand, Lemme *et al.* showed a significant degradation of graphene carrier mobility with more than an 85% decrease for both electrons and holes when an evaporated SiO$_2$ layer was used as a top-gate dielectric.[13]

Here we report the realization of a top-gated graphene field-effect transistor with a high-κ dielectric layer grown by ALD, and with minimal carrier mobility degradation with respect to a graphene layer without a top dielectric. In order to deposit the Al$_2$O$_3$ dielectric, we introduce a thin nucleation layer of oxidized Al between the graphene layer and the dielectric. The electrical characteristics of top-gated field-effect transistors fabricated using this technique, indicate that a high, above 8,000 cm$^2$/Vs, carrier mobility at room temperature after top-gate processing. We develop a simple device model including the effect of quantum capacitance, which agrees well with the observed transport characteristics, and provides the extracted value of mobility, initial charge density, and contact resistance of devices.

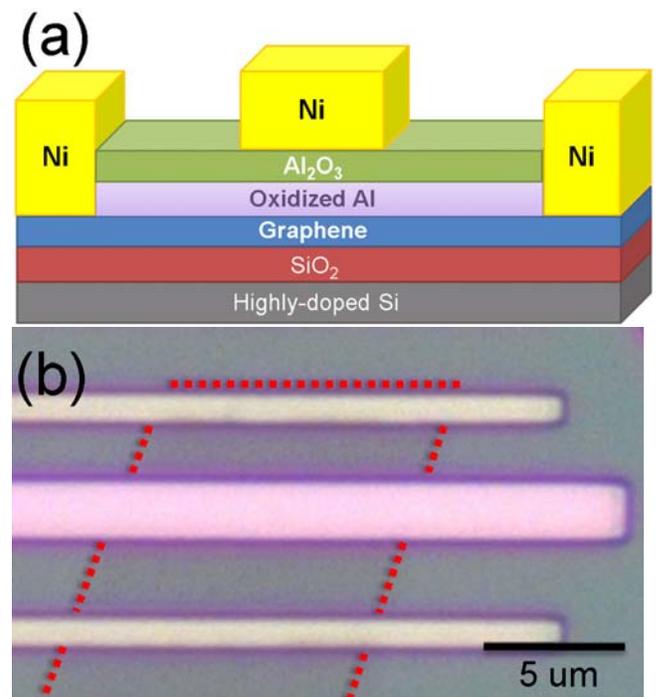

FIG. 1. (Color online) (a) Schematic of dual-gated graphene FET structure. (b) Optical microscope image of a graphene FET.

The key idea enabling the high-κ dielectric layer growth on graphene by ALD is to provide intentional nucleation sites on the inert surface of graphene. Prior to the Al$_2$O$_3$ layer growth by ALD, we deposit a 1-2 nm thick

Al layer on the graphene surface by e-beam evaporation (Fig. 1 (a)). After the Al deposition, the samples are taken out in air and transferred to the ALD chamber for the deposition of $Al_2O_3$, using trimethyl aluminum as the Al source and $H_2O$ as oxidizer. Based on the X-ray photoelectron spectroscopy (XPS) and electrical measurement results, the Al nucleation layer is completely oxidized as soon as the sample is exposed in air to be transferred to ALD chamber[14]. In addition, the initial stage of ALD growth starts with an $H_2O$ oxidizing cycle at elevated temperatures to further complete the oxidation step[15].

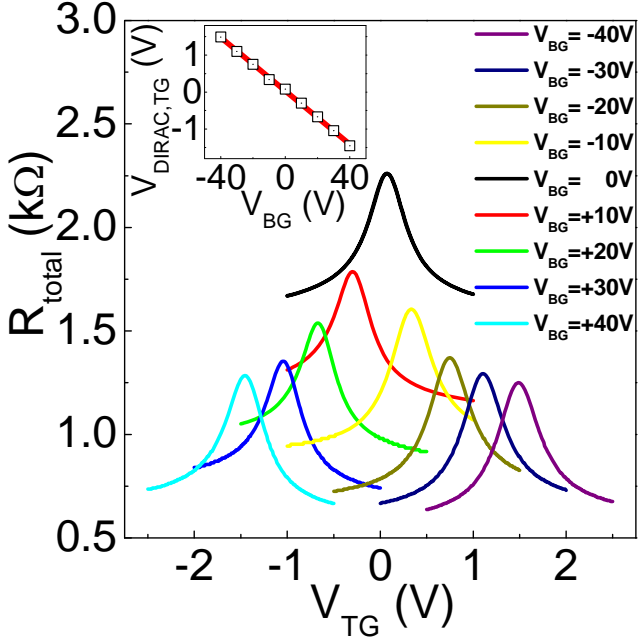

FIG. 2. (Color online) $R_{total}$ vs $V_{TG}$ data measured at different $V_{BG}$ values. The inset shows the position of $V_{DIRAC,TG}$ at different $V_{BG}$.

Graphene monolayer flakes used in this work are exfoliated from bulk natural graphite crystals by the micromechanical cleavage. The substrate consists of a highly-doped, n-type Si (100) wafer with an arsenic doping concentration of $N_D > 10^{20}$ cm$^{-3}$, on which a 300 nm-thick $SiO_2$ layer is grown by thermal oxidation. The low resistivity substrate allows global back-gate operation. The thickness of the exfoliated layers was measured by a combination of optical contrast of the graphene samples[16], thickness measurement by AFM, and Raman spectroscopy[17] to ensure that monolayer flakes are selected for device fabrications. We define metal contacts on the sample using electron beam lithography (EBL) followed by a 50 nm-thick metal (Ni) layer evaporation, and a lift-off process. After annealing in a hydrogen atmosphere at 200°C, which allows the removal of contaminants such as resist residues[18], the device is transferred to a e-beam evaporator vacuum chamber to deposit the Al nucleation layer. Then, the samples are moved to the ALD chamber and go through 167 cycles of $Al_2O_3$ deposition, resulting a 15 nm-thick $Al_2O_3$ film deposition. A 50 nm-thick Ni top-gate electrode is subsequently fabricated using e-beam lithography, metal deposition and lift-off. An example of optical microscope image of a FET, with 6.6-μm source-drain separation and 2.4-μm top-gate length is shown in Fig. 1(b).

The transport characteristics of the device are measured at room temperature in a vacuum probe station. The top-gate electrode and the Si substrate are used as a local gate and global back-gate, respectively, and control the carrier concentration and polarity in the graphene layer. Fig. 2 shows the total device resistance ($R_{total}$) as a function of top-gate voltage measured at different back-gate bias from -40V to 40V, and at a drain bias of $V_D = 0.1V$. Without an applied back-gate bias ($V_{BG} = 0V$) the sample resistance reaches a maximum (Dirac point) at $V_{Dirac,TG} = 0.08V$. This observation indicates that there is little unintentional doping of the graphene sample[19] after the top-gate stack deposition. As |$V_{TG}$-$V_{Dirac,TG}$| increases, the electron or hole concentration in the graphene channel increases and $R_{total}$ decreases, resulting in Λ-shape traces. The top-gate hysteresis is smaller than $0.05V$, and the leakage current through the $Al_2O_3$ top-gate dielectric is less than $0.75\,pA/\mu m^2$. These observations indicate a high dielectric quality and a low ($<9.4\times10^{10}\,cm^{-2}$) interface state density.

Figure 2 data show $R_{total}$ vs. $V_{TG}$ measured at different $V_{BG}$ values. An applied $V_{BG}$ bias changes the position of the Dirac point, and also shifts vertically the measured resistance values. The change of the Dirac point position can be explained as follows: a positive (negative) $V_{BG}$ bias induces a finite concentration of electrons (holes) in the active area, proportional to the back-gate capacitance ($C_{BG}$). In order to restore the device to the Dirac point, where the carrier concentration is minimum, a negative (positive) applied $V_{TG}$ is required. The vertical shift is caused by the resistance change in the un-top-gated regions of the graphene flake. The position of the minimum conductivity points in terms of $V_{TG}$ and $V_{BG}$ is shown in the inset of Fig 2. The slope represents the ratio between the top-gate and back-gate capacitances, $C_{TG}/C_{BG} \approx 28$. Using the back-gate capacitance value of $C_{BG} = 11\,nF/cm^2$, the top-gate capacitance is estimated to be $C_{TG} = 306\,nF/cm^2$, corresponding to a relative dielectric constant of 6.4 for the $Al_2O_3$ film.

We now present a model for the device characteristics of Fig. 2. The carrier concentrations (electrons or holes) in the graphene channel regions, $n_{total}$, can be approximated by

$$n_{total} = \sqrt{n_0^2 + n[V_{TG}^*]^2}$$

where $n_0$ represents the density of carriers at the minimum conductivity, Dirac point. The residual carrier concentration $n_0$, which for an ideal, disorder-free graphene layer should be zero, is generated by charged impurities[20] located either in the dielectric or at the graphene/dielectric interface. $n[V_{TG}^*]$ represents the carrier concentration induced by the top-gate bias away from the Dirac point, $V_{TG}^* = V_{TG} - V_{TG,Dirac}$. The expression for $n[V_{TG}^*]$ is obtained from the following equation relating $V_{TG}$, $C_{ox}$ and the quantum capacitance of the two-dimensional electrons in the graphene channel.

$$V_{TG} - V_{TG,Dirac} = \frac{e}{C_{ox}}n + \frac{\hbar v_F \sqrt{\pi \cdot n}}{e}$$

The total device resistance, $R_{total}$ is given by

$$R_{total} = R_{contact} + R_{channel} = R_{contact} + \frac{N_{sq}}{n_{total}e\mu} = R_{contact} + \frac{N_{sq}}{\sqrt{n_0^2 + n[V_{TG}^*]^2}\,e\mu} \quad (1)$$

where $R_{channel}$ is the resistance of the graphene channel covered by top-gate electrode, and the contact resistance $R_{contact}$ consists of the uncovered graphene section resistance and the metal/graphene contact resistance, and $N_{sq}$ represents the number of squares of the top-gated area.

By fitting this model to the measured data of Fig. 2, we can extract the relevant parameters, $n_0$, $\mu$, and $R_{contact}$. In Fig. 3 we show the measured $R_{total}$ vs. $V_{TG}$ (symbols), along with the model of Eq. (1) (solid lines). The modeling results agree well with the experimental data. Indeed, the data set of Fig. 3 can be fitted with a single value of the residual concentration $n_0 = 2.3 \times 10^{11} cm^{-2}$, of the mobility $\mu = 8,600\, cm^2/V \cdot s$, and with different contact resistance which depend on the applied $V_{BG}$ (Fig. 3 inset).

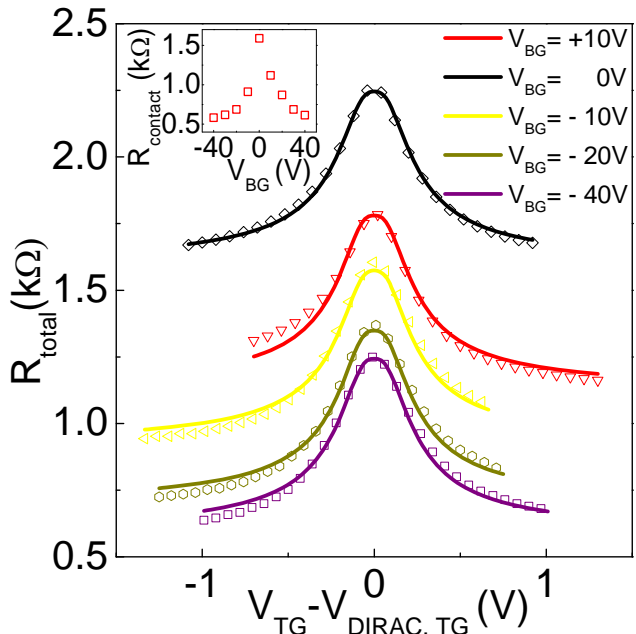

FIG. 3. (Color online) $R_{total}$ vs. $V_{TG}$-$V_{DIRAC,TG}$ at selected $V_{BG}$ values (symbols) along with modeling results for each dataset (lines). The inset shows the extracted contact resistance, $R_{contact}$ vs. $V_{BG}$.

We now discuss the extracted $\mu$ and $n_0$ values in our device in comparison with existing theoretical studies on graphene transport. Adam et al.[20] studied graphene transport in the diffusive limit using the Boltzmann transport formalism, and calculated $\mu$ and $n_0$ as a function of a single parameter, the impurity concentration ($n_{imp}$) at the graphene/dielectric interface[21]: $\mu \cong 33e/(h \cdot n_{imp})$, and $n_0 \approx 0.2 \times n_{imp}$. According to Adam et al.'s model[20,21] the extracted mobility value in our device, $\mu = 8,600\, cm^2/V \cdot s$, corresponds to an impurity concentration $n_{imp} \cong 1.0 \times 10^{12} cm^{-2}$, which in turn would result in a residual carrier concentration $n_0 \approx 1.9 \times 10^{11} cm^{-2}$, in good agreement with our experimental data. Lastly we discuss the temperature dependence of the transport data in our device. From 300K down to 77K the carrier mobility is rather insensitive to temperature, showing a modest ~10% increase. This observation suggests that phonon scattering is relatively small, and that the mobility is primarily determined by fixed impurity scattering[22].

In summary, we fabricated a top-gated monolayer graphene device and successfully deposited an $Al_2O_3$ gate dielectric on its surface by ALD. The device characteristics are investigated in the dual-gate operation mode. Our data show that the overlaying $Al_2O_3$ layer does not substantially degrade the electrical properties of the graphene device. Our model, including quantum capacitance of graphene, agrees very well with our experimental results, and extracted mobility values are above 8,000 $cm^2/V \cdot s$ at room temperature. These results are very promising both for high speed FETs, and also to enable novel device designs in graphene.

We thank D. Yang, R. Ruoff, and S. Adam for useful discussions and support. This work was supported by SWAN-NRI, DARPA and NSF.

[1] K. S. Novoselov, A. K. Geim, S. V. Morozov, D. Jiang, Y. Zhang, S. V. Dubonos, I. V. Grigorieva, and A. A. Firsov, Science **306**, 666 (2004).
[2] K. I. Bolotin, K. J. Sikes, Z. Jiang, G. Fundenberg, J. Hone, P. Kim, and H. L. Stormer, Solid State Commun. **146**, 351 (2008).
[3] T. J. Booth, P. Blake, R. R. Nair, D. Jiang, E. W. Hill, U. Bangert, A. Bleloch, M. Gass, K. S. Novoselov, M. I. Katsnelson, and A. K. Geim, Nano Lett. **8**, 2442 (2008).
[4] B. E. Deal and A. S. Grove, J. Appl. Phys. **36**, 3770 (1965).
[5] J. D. Plummer, M. D. Deal, and P. B. Griffin, *Silicon VLSI Technology Fundamentals, Practice and Modeling* (Prentice–Hall, Upper Saddle River, NJ, 2000).
[6] M. Ritala, K. Kukli, A. Rahtu, P. I. Raisanen, M. Leskela, T. Sajavaara, and J. Keinonen, Science **288**, 319 (2000).
[7] B. Lee, S. Y. Park, H. C. Kim, K. J. Cho, E. M. Vogel, M. J. Kim, R. M. Wallace, and J. Kim, Appl. Phys. Lett. **92**, 203102 (2008).
[8] H. F. Yang and R. T. Yang, Carbon **40**, 437 (2002).
[9] Y. Xuan, Y. Q. Wu, T. Shen, M. Qi, M. A. Capano, J. A. Cooper, and P. D. Ye, Appl. Phys. Lett. **92**, 013101 (2008).
[10] J. R. Williams, L. DiCarlo, and C. M. Marcus, Science **317**, 638 (2007).
[11] X. Wang, S. M. Tabakman, and H. Dai, J. Am. Chem. Soc. **130**, 8152 (2008).
[12] I. Meric, M. Y. Han, A. F. Young, B. Ozyilmaz, P. Kim, and K. L. Shepard, Nat. Nanotechnol. **3**, 654 (2008).
[13] M. C. Lemme, T. J. Echtermeyer, M. Baus, and H. Kurz, IEEE Electron Device Lett. **28**, 282 (2007).
[14] M. J. Dignam, W. R. Fawcett and H. Bohni, J. Electrochem. Soc. **113**, 656 (1966).
[15] C. C. Chang, D. B. Fraser, M. J. Grieco, T. T. Sheng, S. E. Haszko, R. E. Kerwin, R. B. Marcus, and A. K. Sinha, J. Electrochem. Soc. **125**, 787 (1978),
[16] P. Blake, E. W. Hill, A. H. Castro Neto, K. S. Novoselov, D. Jiang, R. Yang, T. J. Booth, A. K. Geim, and E. W. Hill, Appl. Phys. Lett. **91**, 063124 (2007).
[17] A. C. Ferrari, J. C. Meyer, V. Scardaci, C. Casiraghi, M. Lazzeri, F. Mauri, S. Piscanec, D. Jiang, K. S. Novoselov, S. Roth, and A. K. Geim, Phys. Rev. Lett. **97**, 187401 (2006).
[18] M. Ishigami, J. H. Chen, W. G. Cullen, M. S. Fuhrer, and E. D. Williams, Nano Lett. **7**, 1643 (2007).
[19] Y. -W. Tan, Y. Zhang, K. Bolotin, Y. Zhao, S. Adam, E. H. Hwang, S. Das Sarma, H. L. Stormer, and P. Kim, Phys. Rev. Lett. **99**, 246803 (2007).
[20] S. Adam, E. H. Hwang, V. M. Galitski, and S. Das Sarma, Proc. Natl. Acad. Sci. U.S.A. **104**, 18392 (2007).
[21] We use $r_s = e^2/\hbar v_F k = 0.40$ for the coupling constant in our sample; $v_f = 1.1 \times 10^6\, m/s$ is the Fermi velocity in graphene, and $k = 4.98$ is the average dielectric constant of $SiO_2$ and $Al_2O_3$.
[22] J. H. Chen, C. Jang, M. S. Fuhrer, E. D. Williams, and M. Ishigami, Nat. Nanotechnol. **3**, 206 (2008).